\documentclass[preprint2]{aastex}

\slugcomment{Accepted for publication by Astrophysical Journal on June 3rd, 2010}

\begin{document}

\title{A multi-wavelength approach to the properties of Extremely Red Galaxy populations: I - Contribution to the Star Formation Rate density and AGN content.}

\author{H. Messias\altaffilmark{1}, J. Afonso\altaffilmark{1}, A. Hopkins\altaffilmark{2}, B. Mobasher\altaffilmark{3}, T. Dominici\altaffilmark{4} and D.M. Alexander\altaffilmark{5}}
\email{hmessias@oal.ul.pt}

\altaffiltext{1}{Centro de Astronomia e Astrof\'{i}sica da Universidade de Lisboa, Observat\'orio Astron\'omico de Lisboa, Tapada da Ajuda, 1349-018 Lisboa, Portugal.}
\altaffiltext{2}{Anglo-Australian Observatory, P.O. Box 296, Epping, NSW 1710, Australia.}
\altaffiltext{3}{University of California, Riverside, CA 92508, USA.}
\altaffiltext{4}{Laborat\'{o}rio Nacional de Astrof\'{i}sica (MCT/LNA), Rua Estados Unidos, 154, Bairro das Na\c{c}\~{o}es, 37504-364, Itajub\'{a}, MG, Brazil.}
\altaffiltext{5}{Department of Physics, Durham University, South Road, Durham, DH1 3LE, UK.}

\begin{abstract}
\noindent We present a multi-wavelength analysis of the properties of Extremely Red Galaxy (ERG) populations, selected in the GOODS-South/Chandra Deep Field South field. By using all the photometric and spectroscopic information available on large deep samples of Extremely Red Objects (EROs, 645 sources), IRAC EROs (IEROs, 294 sources), and Distant Red Galaxies (DRGs, 350 sources), we derive redshift distributions, identify AGN powered and Star-formation powered galaxies, and, using the radio observations of this field, estimate robust (AGN- and dust-unbiased) Star Formation Rate Densities ($\dot{\rho}_{\ast}$) for these populations. We also investigate the properties of ``pure'' (galaxies that conform to only one of the three ERG criteria considered) and ``combined'' (galaxies that verify all three criteria) sub-populations. Overall, a large number of AGN are identified (up to $\sim 30\%$, based on X-rays, and mid-infrared criteria), the majority of which are type-2 (obscured) objects. Among ERGs with no evidence for AGN activity, we identify sub-populations covering a wide range of average star-formation rates, from below 10\,$M_{\odot}$\,yr$^{-1}$ to as high as 200 $M_{\odot}$ yr$^{-1}$. Applying a redshift separation ($1\leq{z}<2$ and $2\leq{z}\leq3$) we find significant evolution (an increase of a factor of 2 or higher) of $\dot{\rho}_{\ast}$ for EROs and DRGs, while none is observed for IEROs. The former populations can contribute more than 20\% to the global $\dot{\rho}_{\ast}$ at $2\leq{z}\leq3$. The emission from AGN activity is typically not strong in the ERG population, with AGN increasing the average radio luminosity of ERG sub-populations by, nominally, less than 20\%. AGN are common, however, and, if no discrimination is attempted, this could significantly increase the $\dot{\rho}_{\ast}$ estimate (by over 100\% in some cases). Thus, and while the contribution of star forming processes to the radio luminosity in galaxies with AGN remains uncertain, a comprehensive identification of AGN in these populations is necessary to obtain meaningful results.
\end{abstract}


\section{Introduction}

The study of high redshift galaxy populations is fundamental in understanding the formation and evolution of galaxies. One of the major difficulties in such studies is assembling an unbiased and representative sample of galaxies at high-redshift. With the advent of highly sensitive infrared detectors, both ground- and space-based, it has become clear that optical observations reveal only a small and very biased fraction of the galaxy population. This started with the {\it Infrared Astronomical Satellite} ({\it IRAS}) in the 1980s, which found a population of highly luminous galaxies, radiating most of their immense luminosity at infrared wavelengths \citep[1-1000$\,\mu$m,][]{SandersMirabel96}. At about the same time, ground-based telescopes using increasingly more efficient near-infrared (NIR, 1-2.2${\mu}m$) detectors, allowed the detection of very red galaxies \citep[e.g.][]{Elston88,Thompson99}. More recently, observations at (sub-) millimeter wavelengths revealed a population of distant massive galaxies, unseen at optical wavelengths, going through intense (and likely short) episodes of star-formation (SF) activity, possibly leading to the formation of present-day ellipticals \citep[e.g.][]{Barger98,Hughes98}. It is now clear that understanding the $z>1$ universe requires deep multiwavelength observations to build up a complete picture of the processes occurring early in the evolution of galaxies.

In an attempt to constrain hierarchical models of galaxy formation, the last few years have seen optical-to-infrared or infrared-to-infrared color criteria being used to find high-redshift galaxies hosting evolved stellar populations. Extremely Red Objects \citep[EROs,][]{Roche03}, IRAC-selected EROs \citep[IEROs, also known as IR Extremely Red Objects,][]{Yan04} and Distant Red Galaxies \citep[DRGs,][]{Franx03} were thought to identify old passively evolving galaxies at increasing redshifts (from $z>1$ for the EROs/IEROs to $z>2$ for the DRGs), for which a prominent 4000\,\AA\ break would fall between the observed bands. These techniques, however, are also sensitive to active (star-forming, AGN or both) high-redshift dust-obscured galaxies, with intrinsically red spectral energy distributions \citep{Smail02,Alexander02,Afonso03,Papovich06}. These active members of the \textit{Extremely Red Galaxy population} (ERGs, as we will collectively call EROs, IEROs, and DRGs) are also important targets for further study given that they constitute a dusty population of galaxies easily missed at optical wavelengths \citep[e.g.][]{Afonso03}. A challenge in studying the nature of the red galaxy population is the difficulty in disentangling the effects due to redshift, dust-obscuration, and old stellar populations.

Identifying the so-called ``Passively Evolving'' and ``Dusty'' ERGs is a fundamental and particularly difficult task, where optical spectroscopic observations are of limited use. The identification and study of Active Galactic Nuclei (AGN) or star-formation (SF) activity in these galaxies, for example, requires multi-wavelength data from X-ray to radio wavelengths. Radio observations are of particular interest here, given the possibility to reveal the activity in these obscured systems and, for star-forming dominated galaxies, allowing for a dust-free estimate of their SF rates (SFRs).

In this paper we present a comparative study of the ERG population. Using the broad and deep wavelength coverage in the Great Observatories Origin Deep Survey (GOODS)-South/Chandra Deep Field South field, we select samples of EROs, IEROs, and DRGs and estimate their statistical properties. With the extensive photometric data available we explore the redshift distribution, SFRs, and AGN activity in these galaxies. Using radio stacking we estimate dust-free SFRs and the contribution of red galaxy populations with no detected AGN activity to the global SFR density ($\dot{\rho}_{\ast}$).

The structure of the paper is as follows. Sample selection is described in Section \ref{sampsel}. Section \ref{agnid} addresses the AGN identification technique. In Section \ref{ergprop} the ERG sample is characterized, leading to the estimate of the dust- and AGN-unbiased contribution to $\dot{\rho}_{\ast}$. Finally, the conclusions are presented in Section \ref{conc}.

Throughout this paper we use the AB magnitude system\footnote{When necessary the following relations are used:\\($K,~H,~J,~I$)$_{\rm{AB}}$ = ($K,~H,~J,~I$)$_{\rm{Vega}}$ + (1.841, 1.373, 0.904, 0.403) from \citet{Roche03};\\IRAC: ([3.6], [4.5], [5.8], [8.0])$_{\rm{AB}}$ = ([3.6], [4.5], [5.8], [8.0])$_{\rm{Vega}}$ + (2.79, 3.26, 3.73, 4.40) from \emph{http://spider.ipac.caltech.edu/staff/gillian/cal.html}}, and a $\Lambda$CDM cosmology is assumed with H$_{0} = 70$ km s$^{-1}$ Mpc$^{-1}$, $\Omega_{M} = 0.3$, $\Omega_{\Lambda} = 0.7$.

\section{Sample Selection} \label{sampsel}

GOODS was designed to assemble deep multi-wavelength data in two widely separated fields: the Hubble Deep Field North (HDF-N/GOODS-N) and the Chandra Deep Field South (CDF-S/GOODS-S). Specifically the southern field --- GOODS-S --- includes X-ray observations with \textit{Chandra} and \textit{XMM-Newton}; optical ($BVIz$) high resolution imaging with the ACS on board the \textit{Hubble Space Telescope} \textit{HST}; near-infrared (NIR) and mid-infrared (MIR) coverage with the Very Large Telescope (VLT) and the \textit{Spitzer Space Telescope}, respectively; and radio imaging with the Australia Telescope Compact Array (ATCA), Very Large Array (VLA), and Giant Metrewave Radio Telescope (GMRT). These data are among the deepest ever obtained. Large programs aiming at comprehensive spectroscopic coverage of this field are also being performed. The quality and depth of such data make these fields ideal to perform comprehensive studies of distant galaxies and, in particular, of the ERG population.

We use the FIREWORKS $K_s$-band selected catalog from GOODS-South \citep{Wuyts08}. This provides reliable photometry from UV to IR wavelengths (0.2-24${\mu}$m) for each source detected in the $K_{s}$ ISAAC/VLT maps (ISAAC GOODS/ADP v1.5 Release), thus covering an area broadly overlapping the \textit{HST} ACS observations. The widely different resolutions between optical and infrared bands are properly handled to allow consistent color measurements. This is performed by adjusting the optical \textit{HST} and NIR VLT images to a common resolution, and performing photometry on optical, IRAC, and MIPS images, using the prior knowledge about position and extent of sources from the $K_{s}$-band images.

Redshift estimates are also provided. We only use spectroscopic observations to get redshifts with which to derive the intrinsic luminosities of ERGs. Only good spectroscopic redshift determinations \citep[quality flag equal or greater than 0.5 in][]{Wuyts08} were considered, comprising 21\% of the ERGs (Table~\ref{tabctp}). For the remaining sources, photometric redshift estimates from the FIREWORKS catalog were considered. The redshift distributions will be discussed in Section~\ref{reddis}.

The FIREWORKS catalog contains 6308 $K_{s}$ selected sources. To allow for robust selection of our ERG populations we impose a magnitude completeness limit of $K_{s,TOT}=23.8$\,AB (S/N$>3\sigma$) and, following the prescription for robust photometric samples from \citet{Wuyts08}, adopt a pixel weight limit of $K_{s}$w$>0.3$\footnote{See Section 3.4 of \citet{Wuyts08} for a description of the concept of pixel weight.}. This results in a final $K_{s}$-selected catalog of 4740 sources with robust $K_{s}$ photometry.

\subsection{Red Galaxy Samples} \label{selerg}

We consider three categories of ERGs:

\begin{itemize}
\item EROs: $i_{775}-K_{s}>2.48$ \citep{Roche03};
\item IEROs: $z_{850}-[3.6]$\,${\mu}\rm{m}>3.25$ \citep{Yan04};
\item DRGs: $J-K_{s}>1.30$ \citep{Franx03}.
\end{itemize}

Whenever a source is not detected in one of the bands, a limit to its magnitude is assumed (we adopt the $3\sigma$ flux level based on the local rms provided in the catalog). In the case of unreliable photometry (e.g., $K_s$w$\leq0.3$), the corresponding source is not considered further. Thus, only ERGs with robust photometry in both bands (or with a non-detection in the shorter wavelength band) used for their identification are considered.

\begin{figure}
\plotone{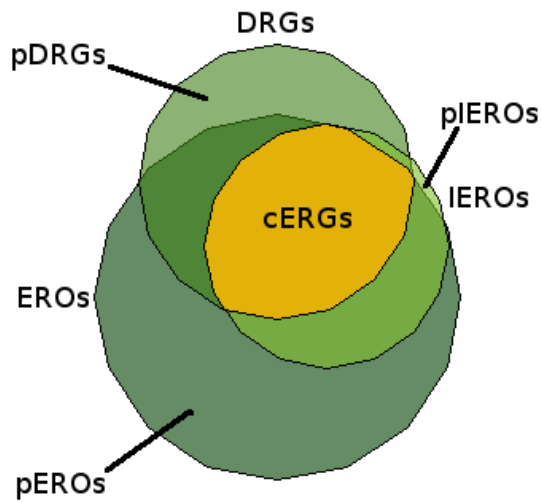}
\caption{Adopted ERG sub-sample nomenclature is represented here through a Venn diagram. The overlap between the three ERG classes -- EROs, IEROs, and DRGs -- is significant (the common ERG population, labeled as cERGs). The outer non-overlapping regions represent the pure populations.\label{venncerg}}
\end{figure}

The resulting ERG sample, with robust photometry, contains 731 objects: 645 EROs, 294 IEROs, and 350 DRGs, down to the adopted magnitude limit of $K_{s,TOT}=23.8$\,AB. These classifications overlap, with individual objects potentially included in more than one classification. This is illustrated in Figure \ref{venncerg}.

It should be noted that FIREWORKS is a $K_{s}$ selected catalog. As such, EROs and DRGs are selected according to the traditional definition, but IEROs selected from the FIREWORKS catalog are in effect $K_{s}$-detected IEROs. This sample will only be representative of the true IERO population in the absence of a significant number of very red $K_s-[3.6]$ IEROs, which are undetected in the $K_{s}$ image. One should also note that the $z_{850}$ detection limit (the current $K_s$ selected sample includes sources with up to $z_{850}\sim$27\,mag) imposes a [3.6]-band magnitude limit of $\sim$23.75\,mag for the IERO sample. 

\begin{figure}
\plotone{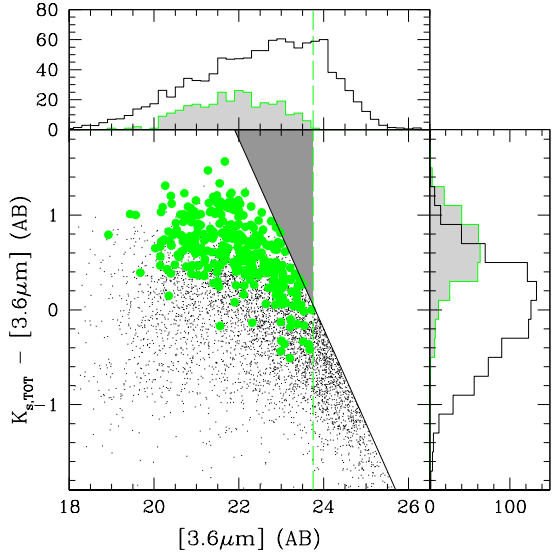}
\caption{$K_s-[3.6]$ color-magnitude plot for sources in the FIREWORKS catalog. Points and open histograms (scaled down by a factor of 4) represent the general $K_s$ population included in the catalog, while filled circles and shaded histograms represent the selected IERO sample. The diagonal line corresponds to a $K_s$-band value of 23.8\,mag, the adopted magnitude limit of our sample. The dashed vertical line represents the practical [3.6]-band IERO magnitude limit (as imposed by the $z_{850}$ 3$\sigma$ magnitude limit and the IERO definition). The current $K_s$-detected IERO sample would differ significantly from the general IERO population if a large number of sources exists above the diagonal and to the left of the vertical line (shaded region). The $K_s-[3.6]$ colors present in the well sampled region of the diagram (below the diagonal line and to the left of the vertical one) argue against this scenario, implying that the current sample of IEROs is representative of the overall IERO population.\label{mciero}}
\end{figure}

Figure~\ref{mciero} shows the color-magnitude distribution for sources in the FIREWORKS catalog and for the $K_{s}$-detected IERO sample, displaying our adopted $K_{s}$-band magnitude limit (diagonal line) and the practical [3.6]-band magnitude limit (vertical line). The sampled region at $K_{s,\rm{TOT}}-[3.6]<-[3.6]+23.8$ and $[3.6]<23.26$ (below the diagonal line and to the left of the vertical one) does not indicate a major incompleteness for the critical region (above the diagonal line and to the left of the vertical one). For example, allowing all FIREWORKS sources to be considered (up to a $K_{s}$-band magnitude of 24.3\,mag), would only increase the IERO sample by 6\% (17 new IEROs). Consequently, we consider our sample of ($K_{s}$-detected) IEROs representative of the true IERO population and find it unnecessary to assemble a separate sample of IEROs from a 3.6\,$\mu$m selected catalog, thus maintaining the photometric homogeneity within the ERG sample.

\subsection{Sub-classes of ERGs} \label{comerg}

We refer to those sources that appear in only one of these classes (ERO, IERO, or DRG) as ``pure'' populations, while those that are simultaneously included in three ERG categories are referred to as the ``common'' population. In this work, the latter will be referred to as ``common'' ERGs, or cERGs. When addressing both the ``pure'' and ``common'' populations we will restrict ourselves to those sources which have sufficient information for a definitive classification in each of the three red galaxy criteria (either good photometry or robust upper limits in \emph{all bands} used for classification). With such requirements, we find 702 ERGs: 626 EROs, 289 IEROs, and 338 DRGs. Essentially all IEROs (290 out of 294\footnote{Number of IEROs with good photometry in the $i_{775}$, $z_{850}$, $K_s$ and 3.6$\mu$m bands.}) are also EROs and more than two-thirds (198 out of 289) are also classified as DRGs; there are 267 sources that comply with both the ERO and DRG criteria, and 197 ERGs that are simultaneously classified as ERO, IERO, and DRG (the cERGs). Identified as ``pure'' sources are 272 pure EROs (pEROs), 3 pure IEROs (pIEROs), and 70 pure DRGs (pDRGs).

Figure~\ref{venncerg} shows the overlap between the different sub-populations. The initial columns of Table~\ref{tabctp} summarize the numbers referred to above.

\section{Multi-wavelength AGN identification and classification} \label{agnid}

One of the major problems for the characterization of ERGs, or for any distant galaxy population, is to identify the presence of AGN activity. The many techniques that exist target different AGN types and redshift ranges, and no single technique can guarantee a high discriminatory success rate. X-rays, radio or MIR, originating from different regions in the vicinity of the AGN, and differently affected by dust obscuration, provide independent ways to reveal such activity. Hence, we use the multi-wavelength data available in this field to carry out a thorough identification and classification of AGN activity in the ERG population.

\subsection{Optical Spectroscopy}

Optical line ratios will only reveal AGN activity if most of the galaxy's line emission comes from the environment near the AGN and if the dust obscuration is not significant. In the case of obscured AGN activity, the emission from any disk SF may dominate the optical line emission. Also, since ERGs are intrinsically UV/optically faint, spectroscopy will be of limited use to reveal their nature. Overall, there are only 11 spectroscopic AGN identifications (narrow line AGN or QSO type-2 classifications), galaxies which are also identified as AGN by the criteria described in the following sections.

Spectroscopy also allows for the rejection of galactic stars selected as EROs. In this sample, 4 were found and discarded from further study.

\subsection{X-Rays} \label{xr}

X-ray emission is arguably the most effective discriminator of AGN activity in a galaxy. Due to the sensitivity levels currently reached with the deepest observations \citep[the 2\,Ms CDF fields: ][]{Alexander03,Luo08}, the most powerful AGN ($L_{0.5-8 keV} > 10^{43}$\,erg\,s$^{-1}$) can be detected beyond the highest redshift currently observed, $z>7$. On the other hand, both low luminosity AGN and vigorous starforming galaxies ($L_{0.5-8 keV} \sim\ 10^{41-42}$\,erg s$^{-1}$) can only be detected out to $z\sim 1-2$. If enough signal is detected, detailed spectral analysis can be used to distinguish between AGN and SF activity as the origin of the X-ray emission.

In this work, the ERG sample was cross-matched with the recent catalogs from the 2\,Ms Chandra observations \citep{Luo08}. For the region considered here -- GOODS-S ISAAC -- the X-ray observations reach aim-point sensitivity limits of $\approx1.9\times 10^{-17}$ and $\approx1.3\times 10^{-16}$\,erg\,cm$^{-2}$\,s$^{-1}$ for the soft (0.5--2.0\,keV) and hard (2--8\,keV) bands, respectively. 

X-ray detections were searched for within 2\arcsec\ of each ERG position. Counterparts were found for 84 of the 645 EROs ($\sim$13\%), 47 of the 294 IEROs ($\sim$16\%), and 51 of the 350 DRGs ($\sim$15\%) (see Table~\ref{tabctp}). These detection fractions are consistent with those found by \citet{Alexander02}, for EROs, and \citet{Papovich06}, for DRGs.

We adopt the X-ray classification criteria from \citet{Szokoly04}, which considers both the X-ray Luminosity ($L_X$), estimated from the 0.5--8\,keV flux, and hardness ratio (HR), calculated using the count rates in the hard band (HB, 2--8\,keV) and in the soft band (SB, 0.5--2\,keV) bands: HR = (HB-SB)/(HB+SB). This is listed as follows:
\begin{eqnarray}
\begin{array}{ll}
{\rm Galaxy:}~ L_{X}<10^{42} {\rm erg\,s}^{-1}~\&~{\rm HR}\leq-0.2 \nonumber \\
{\rm AGN-2:}~10^{41} \leq L_{X}<10^{44} {\rm erg\,s}^{-1}~\&~{\rm HR}>-0.2 \nonumber \\
{\rm AGN-1:}~10^{42} \leq L_{X}<10^{44} {\rm erg\,s}^{-1}~\&~{\rm HR}\leq-0.2 \nonumber \\
{\rm QSO-2:}~L_{X}\geq10^{44} {\rm erg\,s}^{-1}~\&~{\rm HR}>-0.2 \nonumber \\
{\rm QSO-1:}~L_{X}\geq10^{44} {\rm erg\,s}^{-1}~\&~{\rm HR}\leq-0.2 \nonumber 
\end{array}
\end{eqnarray}

The rest-frame X-ray luminosity is calculated as: \[ L_{X} = 4 \pi\,d^{2}_{L}\,f_{X}\,(1+z)^{\Gamma-2}\,\rm{erg}\,\rm{s}^{-1} \] where $d_{L}$ is the luminosity distance in (cm), \emph{f}$_{X}$ is the X-ray flux (erg s$^{-1}$ cm$^{-2}$) in the 0.5-8 keV band and the photon index is assumed to be $\Gamma=1.8$ \citep{Tozzi06}. The luminosity distance is calculated using either the spectroscopic redshift or, if not available, the photometric redshift estimate.

\begin{figure}
\plotone{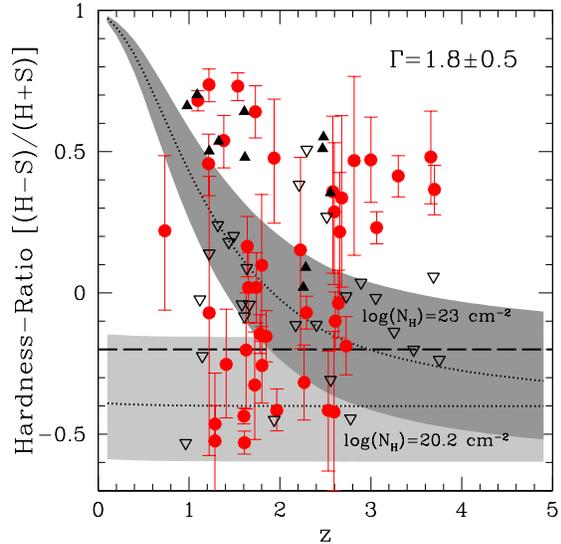}
\caption{X-ray HR evolution with redshift for obscured ($N_H=10^{23}$cm$^{-2}$, gray shaded region) and unobscured ($N_H=10^{20.2}$cm$^{-2}$, light gray shaded region) X-ray PL emission models ($\Gamma=1.8\pm0.5$), calculated using PIMMS (ver. 3.9k). Filled circles show the distribution of the X-ray detected AGN ERGs with a robust HR estimate. Upper limits (no hard-band detection) appear as empty triangles while filled triangles denote lower limits (no soft-band detection). The dashed horizontal line highlights the HR constraint (HR$=-0.2$) for type discrimination used by \citet{Szokoly04}. It is clear that for high-redshift sources ($z\gtrsim2$) the simple HR criterion becomes degenerate as an obscuration measure. \label{hrevol}}
\end{figure}

In total, these criteria enable the identification of 86 sources hosting an AGN with only 6 X-ray sources powerful enough to be classified as QSOs. The majority of the AGN are classified as type-2 sources: 41 X-ray detections have $HR>-0.2$ (i.e., obscured) while only 19 show lower values (with the remaining 26 having uncertain HR determinations, with no discrimination possible), indicating a possible 2:1 obscured to unobscured ratio. However, no constraints on the ratio of type-2 to type-1 AGN can be established, since $HR$ becomes degenerate as an obscuration measure above $z\sim2$, except for the most heavily obscured AGN\citep[$N_{\rm H}\gg10^{23}\,$cm$^{-2}$; see, for example,][]{Alexander05}. Figure~\ref{hrevol} shows the $HR$-redshift evolution for simple obscured and unobscured AGN X-ray emission models, together with the X-ray detected AGN ERGs. Although Type-2 sources seem to dominate (Table~\ref{tabctp}), an X-ray estimate of the Type-2 to Type-1 AGN ratio among such high-redshift populations would require observations extending to lower X-ray energies ($<$0.5 keV) below those reliably achieved by \textit{Chandra}.

\subsection{Mid-Infrared} \label{mir}

Over the last few years with the sensitivity of IRAC and MIPS onboard \textit{Spitzer}, several MIR criteria have been developed for the identification of AGN at the center of galaxies. A power-law (PL) MIR spectral energy distribution (SED), for example, is characteristic of AGN emission \citep[e.g.][]{Donley07}. Somewhat more generic color--color diagrams have also been investigated, and AGN loci in such plots defined \citep[e.g.][]{Ivison04,Lacy04,Stern05,Hatzimi05}. This wavelength range is of particular interest for the ERG population, given their red SEDs. Here, we have applied MIR diagnostics to our ERG sample, as described below.

Observational data at X-ray and IR wavelengths provide complementary views of AGN activity. The most obscured AGN may be missed by even the deepest X-ray surveys but can still be identified by their hot-dust emission at IR wavelengths. On the other hand, depending on the amount of dust and its distribution, and on the AGN strength, the MIR emission from X-ray classified AGN may not be dominated by the hot dust in the vicinity of the AGN itself. A detailed comparison of the merits of AGN selection by the X-rays and the MIR was performed by \citet{Eckart10}, showing that only a multiwavelength combination of AGN criteria can help to overcome biases present in single-band selection. However, even the combination of MIR and X-rays will not result in complete AGN samples, as the identification of low power AGN will ultimately depend on the depth of the surveys. By performing this study in GOODS-S, with some of the deepest data both at X-ray and MIR wavelengths, we maximize the identification rate of AGN.

IRAC counterparts were found for practically all (98\%) ERG sources. The vast majority (92\%) are detected simultaneously in all IRAC bands: 612 of the 645 EROs, 290 of the 294 IEROs, and 322 of the 350 DRGs. The MIPS 24$\mu$m detection rate is understandably lower (55\%), given the lower relative sensitivity: 359/191/222 of the 645/294/350 EROs/IEROs/DRGs are detected (Table~\ref{tabctp}).

\subsubsection {Classification: MIR spectral index} \label{pl}

The IRAC fluxes for each ERG, covering the 3.6 -- 8.0\,$\mu$m range, were fitted with a PL ($f_{\nu}{\propto}\nu^{\alpha}$), classifying as AGN sources those where ${\alpha}<-0.5$ \citep{Donley07}. To increase its reliability, application of this technique was limited to sources simultaneously detected in all of the four IRAC bands, and the $\alpha$ value was only accepted if the $\chi^2$ probability fit was $P_{\chi^2}>0.1$.

This criterion reveals the existence of AGN in 6 (1\%) EROs, 1 of which is also identified as an AGN in X-rays; in 4 (1\%) of the IEROs, 1 of which have an X-ray AGN classification; and in 12 (3\%) of the DRGs, none having an X-ray AGN classification (see Table~\ref{tabctp}).

\subsubsection {Classification: MIR colors} \label{st}

In recent years, several AGN color-selection criteria have been developed employing MIR IRAC observations \citep{Ivison04,Lacy04,Stern05,Hatzimi05}. They are not completely independent from the PL criteria mentioned above, but are likely more sensitive to other kinds of AGN \citep[e.g., QSOs or broad-line AGN,][]{Lacy04,Stern05}. Here we follow the criterion proposed by \citet{Stern05}, which identifies AGN from galaxy populations if they satisfy the following relations:
\begin{eqnarray}
\begin{array}{l}
([5.8] - [8.0]) > -0.07 \nonumber \\
([3.6] - [4.5]) > 0.2([5.8] - [8.0]) - 0.156 \nonumber \\
([3.6] - [4.5]) > 2.5([5.8] - [8.0]) - 2.295 \nonumber
\end{array}
\end{eqnarray}

\begin{figure}
\plotone{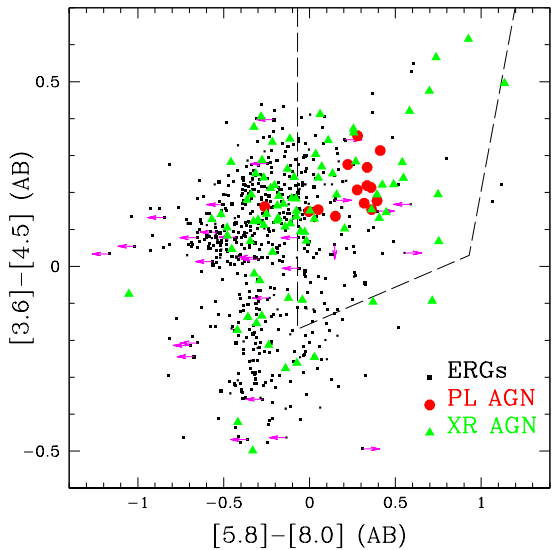}
\caption{Distribution of ERGs on the color--color diagnostic plot proposed by \citet{Stern05} with the AGN region delimited by the dashed line. The ERGs classified as AGN by the PL \citep{Donley07} and X-rays \citep{Szokoly04} criteria are highlighted (red circles and green triangles, respectively).\label{stern}}
\end{figure}

Figure~\ref{stern} shows the distribution of ERGs on the color--color plot of \citet{Stern05}. It identifies as AGN 135 (21\%) EROs, 27 of which are also classified as AGN from the X-rays; 83 (28\%) IEROs, 21 of which also have an X-ray AGN classification; and 117 (33\%) DRGs, 25 of which also appear as X-ray AGN. The relatively high number of potential AGN identified, over that revealed by the X-rays, is known and expected \citep[][and references therein]{Donley07}.

\subsubsection{MIR degeneracy at $z>2.5$} \label{mir25}

One problem in using MIR photometry to identify AGN (with both PL and color--color criteria) arises at $z \gtrsim 2.5$, as both star-forming galaxies and AGN start to merge into the same MIR color--color space. The main reason for this is the increasing relative strength of stellar emission in the MIR, as compared to that of an AGN, as redshift increases. At higher redshifts, a prominent 1.6\,$\mu$m stellar bump may pass through the IRAC bands, allowing for the detection of a steep spectral index not from AGN emission, but from the stellar emission alone.

At z$>$2.5, the MIR spectral index method (see Section \ref{pl}) identifies AGN in 3 EROs, 3 IEROs, and 10 DRGs, where none is identified through X-ray emission; on the other hand, the color--color method of \citet{Stern05} classifies as AGN 51 EROs, 42 IEROs, and 76 DRGs, with 14, 11, and 16 (respectively) X-ray confirmed at these redshifts.

In the present work, this is not a serious problem, as most of the ERG sample (77\%) lies at $z\leq2.5$ (see Section \ref{reddis}). Nevertheless, it should be noted that at higher redshifts, this could result in a likely overestimate of the presence of AGN. One can attempt to correct for this effect, by using the MIPS 24$\mu$m observations: at $z\sim2.5-5$, the 1.6\,$\mu$m bump will be shifted to the 6--10\,$\mu$m range. Therefore, in the absence of significant AGN emission, one expects a blue [8.0]-[24] color. We have investigated the [8.0]-[24] colors of $z>2.5$ ERGs, removing a MIR AGN classification whenever [8.0]-[24] $<$ 1. In this case, the observed MIR emission does not require an AGN component, but just a more or less evolved stellar population.

\begin{figure}
\plotone{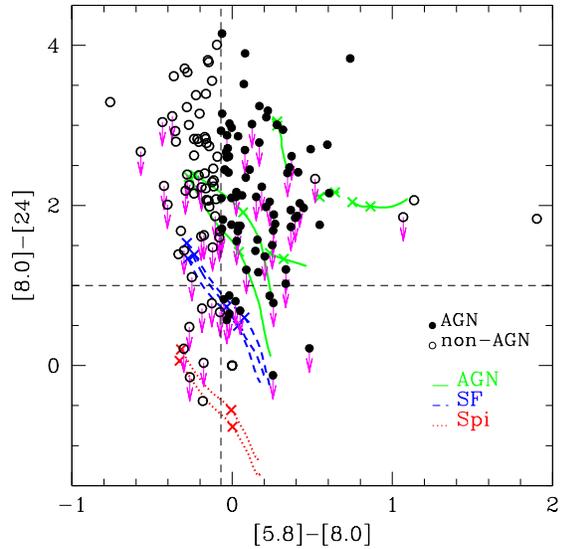}
\caption{Mid-infrared [5.8]-[8.0] versus [8.0]-[24] color--color plot for $z>2.5$ ERGs in the current sample. Filled symbols represent an AGN classification from a MIR indicator (IRAC PL or IRAC color--color) and open symbols otherwise. The tracks represent the expected colors of template SEDs where the IR is dominated by star-formation (dotted and dashed tracks, where the latter represent more intense SF activity) or AGN activity (continuous tracks), redshifted between $z=2.5$ and $z=4$, with crosses at $z=2.5$ and $z=3$. The templates displayed are (from bottom to top): 2 Spirals (Sc and Sd), 3 starbursts (M82, NGC 6240, and Arp220), and 6 AGN (IRAS 22491-1808, IRAS 20551-4250, QSO-2, \citet{Afonso01} ERO, Mrk 231, and IRAS 19254-7245 South).\label{z3agn}}
\end{figure}

In Figure ~\ref{z3agn} we present the MIR [5.8]-[8.0] versus [8.0]-[24] color--color plot for $z>2.5$ ERGs in the current sample (highlighting those classified as AGN by the two MIR criteria referred to in Section \ref{pl} and Section \ref{st}). The tracks represent the expected colors of template SEDs when redshifted between $z=2.5$ and $z=4$. All templates were taken from the SWIRE Template Library \citep{Polletta07}, with the exception of the track in the upper right, corresponding to the extreme ERO of \citet{Afonso01}, the MIR emission of which is dominated by an obscured AGN. The vertical line indicates the [5.8]-[8.0] color constraint of the \citet{Stern05} criterion, while the horizontal line shows our adopted color cut separating AGN and star-forming processes at these redshifts. The AGN template that crosses over this [8.0]-[24] threshold at the highest redshifts is IRAS 22491-1808, a mixture of AGN and stellar MIR emission, where the AGN component is progressively less sampled by the MIR bands as redshift increases. Concerning the current sample, the few high redshift ($z>2.5$) sources classified as AGN by the MIR criteria that appear below the [8.0]-[24] threshold, 10/7/10 EROs/IEROs/DRGs, {\it do not require} an AGN SED to explain their MIR emission and, consequently, their MIR AGN classification is removed. We note that the MIPS (24$\mu$m) - IRAC colors have not yet been considered in detail as a way of selecting AGN/SF sources in deep MIR surveys \citep[however, see][]{Ivison04,Lacy04}. Prompted by Figure ~\ref{z3agn} we are currently performing a detailed analysis of such colors as potential diagnostics for AGN behavior at high-z, to be presented in a subsequent paper.

We note the presence of two interesting sources in Figure~\ref{z3agn}. The one isolated in the upper right, is one of the seven optically unidentified radio sources found in \citet[][their source \#42]{Afonso06}. Inspection of the $K_s$ and 24$\mu$m images reveals no signs of blending, strengthening the accuracy of the 24$\mu$m flux. This source also has X-ray emission characteristic of an AGN ($L_{X}=10^{43.3} {\rm erg\,s}^{-1}$). The FIREWORKS assigned photometric redshift is $z=3.69$ (in agreement with our own detailed ongoing investigation of this source, resulting in an estimate of $z=3.85$ using Hyper--$z$). The high-$z$ obscured AGN scenario postulated in \citet{Afonso06} for this source is thus strengthened. The color-track closest to this source in Figure~\ref{z3agn} is that of the highly obscured AGN ERO found by \citet{Afonso01}.

The other interesting source is the bluest [8.0]-[24] 24$\mu$m detection, with MIR colors characteristic of spiral galaxies. It is also X-ray detected but has no radio emission. This is a candidate for a high-$z$ evolved system with MIR colors typical of Spiral c type. The redshift assigned to this source, $z_{\rm{phot}}=2.53$, is at the upper limit of the redshift range in which these type of sources have ever been found \citep{Stockton08}. A few more galaxies fall in the same region of the color--color plot. Again they are candidates for Spiral type evolved systems whose study is relevant to constrain hierarchical models for the formation of galaxies.

\subsection{Radio}

Radio emission is essentially unaffected by dust obscuration, making it a highly desirable diagnostic for SF activity in ERGs. Since both SF and AGN activity can produce radio emission, it is often difficult or impossible to rely on radio properties alone to reveal the power source in a galaxy. Indications from radio spectral indices are of limited use, as both SF and AGN emission usually result from synchrotron radiation with $S_\nu \propto \nu^{-0.8}$, and only some AGN show signs of flat or even inverted radio spectra. Very high resolution VLBI radio imaging has also been used with limited success to impose limits on the size of the radio emitting region, identifying star-forming galaxies where the radio emission is resolved, and a possible AGN where not \citep{Muxlow05,Middelberg08,Seymour08}. The only straightforward radio AGN criterion is the radio luminosity itself, as the highest luminosities can only be produced by the most powerful AGN.

\citet{Afonso05} performed a detailed study of the sub-mJy radio population, and found star-forming galaxies with radio luminosities up to $L_{\rm 1.4\,GHz} \sim 10^{24.5}\,$W\,Hz$^{-1}$. We thus take this value as the upper limit for SF activity. We note that this value corresponds to an SFR of almost 2000\,M$_{\sun}$\,yr$^{-1}$ \citep[][see Section \ref{sfrfin}]{Bell03}. The existence of galaxies with higher rates of SF activity is unlikely. 

For the current work we have used the 1.4\,GHz ATCA observations of this field, which reach a uniform 14--17\,$\mu$Jy rms throughout the GOODS-S field \citep[see][for more details]{Afonso06,Norris06}.

We used as 3\arcsec matching radius to search for radio emission from the ERGs. In total 17 ERGs have reliable radio detections ($\gtrsim 4.5 \sigma$): 15 EROs (2.5\% of this population), 10 IEROs (3.3\%), and 11 DRGs (3.1\%). Only 3 of these sources have radio luminosities in excess of $10^{24.5}\,$W\,Hz$^{-1}$. They are also classified as AGN by the previous X-ray and MIR criteria. On the other hand, only 3 radio-detected ERGs are not classified as AGN by any of the adopted criteria. For these sources, the radio luminosity corresponds to SFRs of 93, 646, and 1584 $M_{\odot}$ yr$^{-1}$ \citep[using the conversion from][see Section \ref{sfrfin}]{Bell03}.

The small detection rate indicates that powerful AGN and the most intense starbursts are not common in the ERG population, as only sources with L$_{1.4\rm{GHz}}>10^{23}$\,W\,Hz$^{-1}$ will be detected at $z>1$ with the sensitivity available even in the current deepest radio surveys.

\section{Properties of ERGs} \label{ergprop}

\subsection{Redshift Distributions} \label{reddis}

As noted above, robust spectroscopic redshifts are available for around 21\% of the ERG sample. Photometric redshift estimates are also available from the FIREWORKS catalog, covering almost the complete ERG sample.

\begin{figure}
\plotone{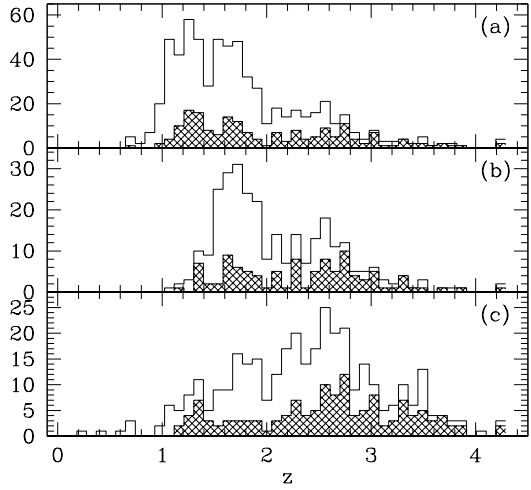}
\caption{Redshift distributions of different ERG sub-populations: (a) EROs, (b) IEROs, and (c) DRGs. The hatched histograms correspond to ERGs identified as AGN. Note the different $y$-axis scales for the individual panels.\label{geral}}
\end{figure}

The redshift distributions for the ERO, IERO, and DRGs are shown in Figure~\ref{geral}. Although the range of redshifts sampled in all ERG classes is similar, the average value increases from $z=1.76$ for EROs, to $z=2.11$ for IEROs, and to $z=2.40$ for DRGs populations \citep[in agreement with previous works, e.g.][]{Conselice08,Papovich06}. This was expected given the source selection, designed to identify objects at such redshifts. 

The AGN in the ERG population follow a similar redshift distribution but the AGN fraction increases rapidly at higher redshifts. This will be addressed in the next section.

\begin{figure}
\plotone{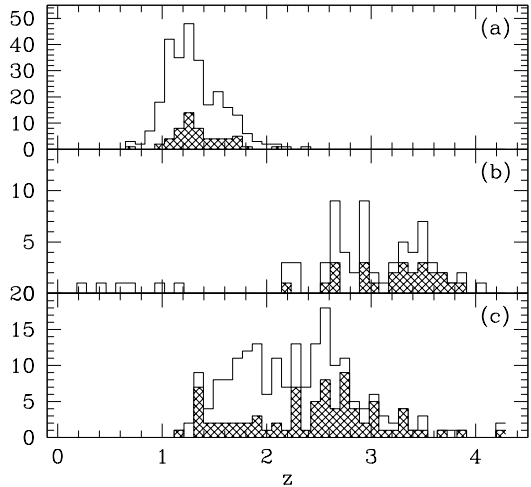}
\caption{Redshift distributions for ``pure'' and ``common'' ERG sub-populations: (a) pEROs, (b) pDRGs and (c) cERGs. The hatched histograms correspond to ERGs identified as AGN. Note the different $y$-axis scales for the individual panels.\label{purecom}}
\end{figure}

Figure~\ref{purecom} displays the redshift distributions once again but for pEROs, pDRGs, and cERGs (only 3 pIEROs exist, at redshifts $z_{\rm{phot}}=1.7, 3.0$ and $z_{spect}=2.6$). The redshift distribution of pEROs is quite narrow, selecting sources essentially at $z=1-2$ (peaking at $z\sim 1.3$), while the pDRG population is notably small, and at higher redshifts ($z=2-4$). The ``pure'' criteria thus appear to be good and easy techniques to select high-z sources in narrow distinct redshift bins. Sources classified as cERGs, appearing as red in all three ERG selection criteria, cover a broad redshift range, from $z=1$ to $z=4$. There are no cERGs at $z<1$, in this particular sample due to the IERO criterion.

\subsection{AGN content of ERGs} \label{agncont}

As described in the previous section, several multi-wavelength indicators were used to identify AGN in the ERG population. The indicators have different sensitivities to AGN characteristics, such as distance, dust obscuration, or AGN strength. Their combination will, thus, allow for a more complete census of AGN content in these sources.

We do not find a numerous population of very powerful AGN among the ERGs, as given by the X-rays (L$_{X}\geq10^{44}$\,erg s$^{-1}$, 6 ERGs) and radio (L$_{\rm 1.4\,GHz}\geq 10^{24.5}\,$W\,Hz$^{-1}$, 3 ERGs) luminosities. These represent, respectively, only 0.8\% and 0.2\% of the ERG sample, comparable to the fraction observed in the complete $K$-selected FIREWORKS sample, where 17 (0.4\%) QSOs and 5 (0.1\%) radio-powerful sources are found. Overall, we select 214 (29\%) AGN-dominated systems in the ERG sample (29\% for EROs, 35\% for IEROs, and 39\% for DRGs). This fraction increases from low to high redshift, from 22\% at $1\leq{z}<2$ to 38\% at $2\leq{z}\leq3$. Among the X-ray identified AGN, 35\% are also classified as such by the Stern `wedge'. Conversely, only 18\% of the `wedge' identified AGN are X-ray detected. All but one of the PL classified AGN fall inside the Stern `wedge' (Figure~\ref{stern}), as expected from the non-independence of both indicators \citep[as can be seen in the detailed comparison by][]{Donley08}.

The high AGN fraction and its increase with redshift, might lead one to think that the MIR criteria are overestimating the number of AGN at high redshifts, even though a tentative correction was applied (see Section \ref{mir25}). We have investigated the AGN fraction evolution from $1\leq{z}<2$ to $2\leq{z}\leq3$ based, independently, on the X-ray and MIR indicators. In both wavebands, the AGN fraction increases significantly from low to high redshifts, rising from 9\% to 16\% when the X-ray is\ considered and from 15\% to 29\% when the MIR is considered. Although it seems possible that SF galaxies may still be affecting the MIR criteria at high redshift (see Section \ref{mir25}), most of the AGN fraction increase is consistent between the two wavelength indicators. This may partly be an effect of Malmquist bias, with lower luminosity systems, more likely to be dominated by SF, being progressively lost at higher redshift. In any case, this increase is consistent with the known history of AGN activity in the universe \citep{Shaver96,Hopkins07}.

\begin{figure}
\plotone{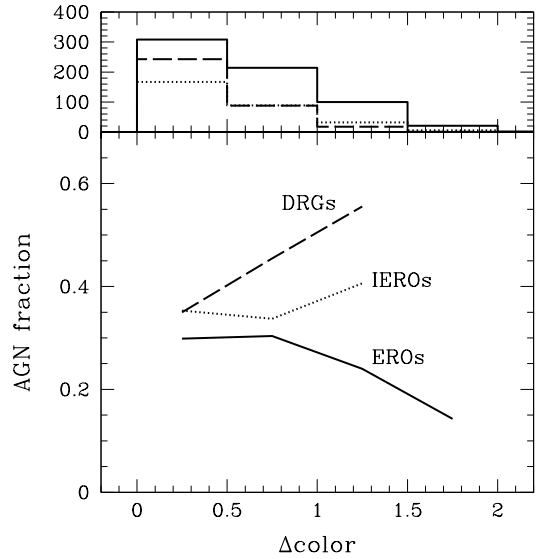}
\caption{AGN fraction as a function of color for EROs, IEROs, and DRGs. The x-axis represents the difference in color to the color threshold adequate for each population: $i_{775}-K_s=2.48$ for EROs, $z_{850}-[3.6]=3.25$ for IEROs and $J-K_s=1.3$ for DRGs. IEROs appear to have an intermediate behavior relative to EROs and DRGs, as one goes to more extreme colors. The histograms on the upper part of the picture show the color distribution for each ERG sub-population.\label{coragn}}
\end{figure}

\begin{figure}
\plotone{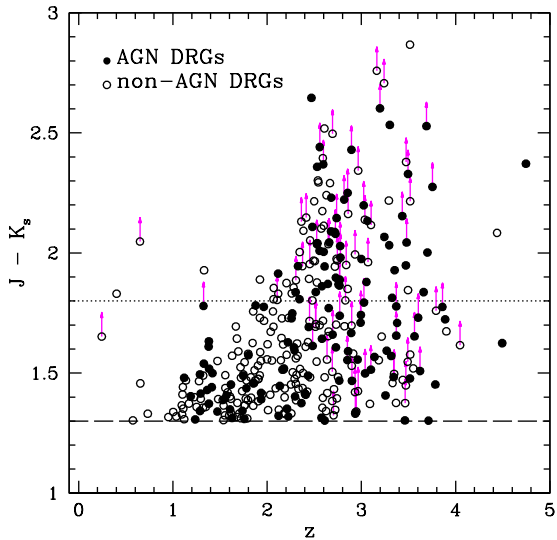}
\caption{Variation of $J-K_s$ color with redshift. The DRG criterion color cut is shown as a dashed line. A 0.5\,mag redder $J-K_s$ cut (dotted line) selects almost no $z<2$ DRGs.\label{jkvsz}}
\end{figure}

The AGN fraction is not a simple function of color, as shown in Figure~\ref{coragn}. All three sub-populations (EROs, IEROs, and DRGs) have similar AGN fractions around the color threshold ($\sim30\%$), but show different behavior with increasing colors. The more extreme colors among the ERGs do not necessarily correspond to higher fraction of AGN identifications, and in fact, the opposite appears to be true for EROs. The AGN fraction is high only for the more ``moderate'' $i-K_s$ colors, decreasing for $i-K_s\gtrsim3.7$. The difference between the ERO and DRG trends result from each criterion itself. Redder $i-K_s$ colors will always select a population mostly at low-$z$ ($1\leq{z}<2$), where the AGN fraction is shown to be smaller. On the other hand, redder $J-K_s$ constraints imply a higher fraction of high-$z$ ($2\leq{z}\leq3$) sources, where the AGN fraction is higher. For example, essentially no low-$z$ source has $J-K_s\gtrsim1.8$ (Figure~\ref{jkvsz}). The AGN fraction versus color trend of the IEROs lies between that of the EROs and DRGs, which could be driven by the redshift distribution of the IERO population, which also lies between that of the EROs and DRGs.

\subsection{Radio Stacking} \label{stack}

Another important aspect necessary in understanding the properties of ERGs is their SFR and the contribution of these populations to the overall SFR density of the universe. Dust obscuration is a serious source of uncertainty in estimating SFRs in ERGs from rest-frame ultraviolet luminosities. A SF diagnostic not affected by dust obscuration is radio emission. However, these galaxies are distant enough that even the deepest radio surveys are sensitive to only the brightest star-forming population (detection limits corresponding to several hundred $M_{\odot}$\,yr$^{-1}$ for $z\gtrsim 1$). Instead, stacking methods can be used to evaluate the statistical star-forming properties of ERGs. Stacking, as used here, is simply an ``image stacking'' procedure, where image sections centered at each desired source position (stamps) are combined. The aim is to reach much lower noise levels, possibly providing a statistical detection of samples whose elements are individually undetected in the original image.

For the radio stacking analysis we have used the 1.4\,GHz ATCA observations of this field, reaching a uniform 14--17\,$\mu$Jy rms throughout the GOODS-S field \citep[see][for more details]{Afonso06,Norris06}. Our adopted stacking methodology can be summarized in the following steps. 

First, using the radio image of the field, stamps of 60 by 60 pixels (equivalent to 120'' by 120'') were deemed appropriate, allowing a good sampling of the vicinity of each source, necessary to identify strong neighboring sources that can bias the stacking.

Every stamp containing a radio source within an 18$\arcsec$ radius from the central (ERG) position, was rejected. As the stacking will be used to estimate the average flux of the unidentified ERGs in the radio image, the inclusion of radio detections would likely bias the final result. In this context the term ``detection'' does not only apply to the robust detections (roughly at $>4.5\sigma$ level), but also to the ``possible'' detections (all remaining candidate radio sources at $>3\sigma$). Stamps having radio sources near the ERG position must also be excluded, as the wings of the radio detection can extend to the central part of the stamp.

The remaining stamps for each sample of ERGs can then be stacked. Previous work often uses median stacking \citep[e.g., ][]{White07}, in an attempt to be robust to radio detections and high/low pixels. The penalty for this is the loss of sensitivity. Having removed all detections and possible detections from the list of stamps, we use a weighted average (weight=rms$^{-2}$) stacking procedure. At each pixel position we further implement a rejection for outliers, rejecting high (low) pixels above (below) the $3\sigma$ ($-3\sigma$) value {\it for that pixel position}. The number of rejected pixels in the central region is always very low ($\leq2$) confirming that previous rejection steps work efficiently. 

The final flux and the noise level are measured in the resulting stacked image. To evaluate the reliability of detections in the stacked images we performed Monte Carlo (MC) simulations. Random positions in the radio image were selected and stacked, following the procedure described above. Each of these positions were required to be farther than 6\arcsec from the known $K_s$ sources, as we are interested in evaluating systematics of the radio image alone. Appropriate numbers of stacked stamps were used, to compare to the actual ERG stacks. The procedure was repeated 10000 times for a given number of stamps. For further reference, a stacked sample will be considered to have produced a reliable detection only if no MC simulation (among 10,000) has resulted in a higher S/N value.

\subsection{Star formation activity in ERGs} \label{sfrfin}

Following the procedure outlined above, we have performed a radio stacking analysis for different sub-groups within the ERG population. The radio data were stacked for each of the populations of EROs, IEROs, DRGs, pEROs, pDRGs, and cERGs. Within these samples, stacking of the radio images was also performed separately for AGN and non-AGN sub-populations. Since redshift estimates exist for the vast majority of the ERGs, stacking is performed separately for both low and high redshifts ($1\leqslant{z}<2$ and $2\leqslant{z}\leqslant$3, respectively). Besides minimizing biases in the stacking signal, due to different populations and different (radio) luminosities being sampled at different redshifts, this also allows us to search for a hint of any evolutionary trend. Given the incompleteness of the sample at the highest redshifts no attempt was made to perform a specific radio stacking analysis for $z>3$ ERGs. Table~\ref{tabnum} lists the number of sources considered in each of the sub-populations and those in each of the stacking steps referred in the previous section.

\begin{figure}
\plotone{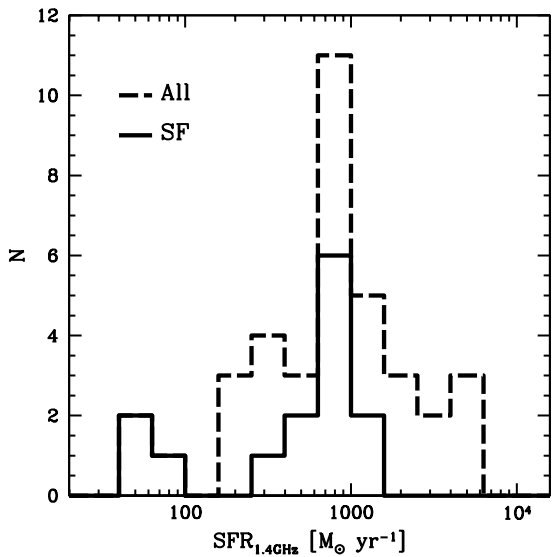}
\caption{SFR distribution of the radio detected ERGs (considering any signal in the radio map with $>3\sigma$). Dashed histograms show the overall distribution, while continuous histograms refer to the sources considered as SF systems.\label{sfrdist}}
\end{figure}

While the stacking procedure enables the average flux to be estimated from the radio-undetected sample ($<3\sigma$ signal), the entire population should be considered when measuring the ERG contribution to the global SFR density of the universe. The approach adopted here was to consider all radio-undetected ERGs as having a radio flux given by the average signal from the stacking analysis, and all radio-detected ERGs\footnote{For this purpose, radio-detections refer to signals above $3\sigma$ in the radio map; see Section \ref{stack}.} to contribute with their measured flux density. The conversion from radio flux to radio luminosity is performed by using the assumed redshift (spectroscopic or photometric) and a radio spectral index of $\alpha=0.8$ ($S_\nu \propto \nu^{-\alpha}$): \[ L_{1.4\rm{GHz}} = 4 \pi\,d^{2}_{L}\,S_{1.4\rm{GHz}}\,10^{-33}(1+z)^{\alpha-1}\,\rm{W}\,\rm{Hz}^{-1}, \] where $d_{L}$ is the luminosity distance (cm) and $S_{1.4\rm{GHz}}$ is the 1.4 GHz flux density (mJy). The corresponding SFR is obtained using the calibration from \citet{Bell03}:
\begin{eqnarray}
\rm{SFR}~(\rm{M}_{\odot}~\rm{yr}^{-1})=\left\lbrace
\begin{array}{lcc}
5.52\times10^{-22}L_{1.4\rm{GHz}} &,&L>L_c \nonumber \\
\frac{5.52\times10^{-22}}{0.1+0.9\left(\frac{L}{L_c}\right)^{0.3}}L_{1.4\rm{GHz}} &,&L{\leq}L_c \nonumber
\end{array} \right.
\end{eqnarray}
where $L_c=6.4\times10^{21}$ W Hz$^{-1}=10^{21.81}$ W Hz$^{-1}$. The contribution to $\dot{\rho}_{\ast}$ was estimated for individual galaxies using the $1/V_{\rm{max}}$ method \citep{Schmidt68}. The associated volume for each galaxy is estimated by using a $K$-correction derived from the galaxy's own SED (as given by the observed multi-wavelength photometry). Again, radio-detected ERGs, contributed with their estimated intrinsic luminosity and SFR, derived with the assigned redshift estimate and its detected flux. In Figure~\ref{sfrdist}, the SFR distribution of these sources is presented. Those classified as SF (14 in total), range from $\sim$50 to $\sim$1600 $M_{\odot}$ yr$^{-1}$ \citep[in reasonable agreement with those presented in][]{Georgakakis06}. On the other hand, the luminosity and SFR estimates of radio-undetected ERGs were based on the resulting stacking signal of the sample and, likewise, the individual ERG redshift value.

The results are given in Table~\ref{tabsta}. For each ERG sub-population we list: (1) the ERG sub-population; (2) the total number of sources in the sample; (3) the final number of stamps included in the stacking; (4) the rms of the final stacked image; (5) the measured flux in the central region of the stacked image; (6) the respective S/N; (7) number of Monte-Carlo simulations (out of 10000) that resulted in higher S/N values, a measure of the reliability of the ERG detection: conservatively, whenever $N_{MC}>0$ the stacking signal is considered spurious; (8) the average redshift for the sub-population; (9) the average radio luminosity both for the radio non-detected sources -- taking into account the stacking signal only -- and, in parenthesis, that for the entire sub-population (including radio detected sources); (10) the average SFR, for non-AGN samples, corresponding to the radio luminosities in column (9); (11) the resulting radio luminosity density ($\mathcal{L}_{1.4\,{\rm GHz}}$); (12) the corresponding $\dot{\rho}_{\ast}$. For columns (9) to (12), the upper limits, corresponding to non-detections of the stacked signal ($N_{\rm{MC}}>0$), are estimated using the maximum S/N found on the MC simulations. No stacking is attempted for populations with less than 10 stamps.

The analysis suggests that the bulk of the ERO population\footnote{In this section, while discussing the SF properties of ERGs, we will naturally be referring to the non-AGN sources, dropping the prefix 'n' for simplicity.} have modest SF activity. At $1~{\leq}~z<2$, where most EROs are found, the average SFR is below a few $M_{\odot}$ yr$^{-1}$. Only at $2~{\leq}~z\leq3$, EROs -- practically all (93\%) being classified as DRGs -- reveal intense average SFRs, $\sim200$ $M_{\odot}$ yr$^{-1}$, entering the Ultra Luminous Infrared Galaxies (ULIRG) regime. This suggests that at low-$z$ the passive/evolved systems represent a significant fraction of the ERO population (56\%, see pEROs discussion ahead), as opposed to the high-$z$ regime where the dusty systems dominate. DRGs show the same trend although not as pronounced: they increase from $\sim70$ $M_{\odot}$ yr$^{-1}$ at $1~{\leq}~z<2$ to $\sim140$ $M_{\odot}$ yr$^{-1}$ at $2~{\leq}~z\leq3$. Interestingly, at low redshifts ($1~{\leq}~z<2$), 97\% of DRGs (69 out of 71) are also classified as EROs; conversely only 20\% of the low-$z$ EROs (69 out of 344) are also classified as DRGs. The estimated average SFR of $\sim70$ $M_{\odot}$ yr$^{-1}$ for this overlapping population supports previous claims of a dusty starburst nature for these sources \citep{Smail02,Papovich06}.

These values for the DRG population are comparable to those found in the literature \citep{Rubin04,Forster04,Knudsen05,Reddy05}. \citet{Papovich06} studied 153 DRGs selected also in the GOODS-S to a limiting magnitude of $K_{s,\rm{TOT}}<23$AB. They find an average SFR for the DRG population at $1\lesssim{z}\lesssim3$ of $200-400$ $M_{\odot}$ yr$^{-1}$, which is somewhat higher than our result. This difference is attributable to the different methods of DRG selection. First, the DRG selection in \citet{Papovich06} is slightly redder (0.06 mag) and restricted to brighter $K$-band magnitudes (0.6 mag difference). The first alone would be responsible for an increase of $\sim$10\% in our estimated SFR, while the latter effectively restricts the sample to stronger starbursts, increasing the average SFR by $\sim$30\%: from 100 $M_{\odot}$ yr$^{-1}$ at $K_{s,TOT}<23.8$ to 130 $M_{\odot}$ yr$^{-1}$ at $K_{s,TOT}<23.0$\citep[see also ][]{Reddy05}. Second, we employ a more conservative AGN rejection, by using the full 2Ms \textit{Chandra} data and rejecting essentially all sources classified as AGN from MIR color--color diagrams. \citet{Papovich06} look at the individual MIR IRAC bands to re-classify as starforming a considerable number of DRGs flagged as AGN by \citet{Stern05}. Finally, the uncertainty in the different methods and associated calibrations (UV+IR luminosities versus Radio luminosities) used to estimate the SFR account for the remaining observed difference.

The low average SFR for EROs at $1~{\leq}~z<2$ is due to the numerous pEROs (193, 56\% of the $1~{\leq}~z<2$ EROs): the stacking analysis of pEROs found in this redshift range fails to produce any signal. This population likely corresponds to the passively evolving component of EROs. On the other hand, pDRGs at $2~{\leq}~z\leq3$ must also be characterized by relatively low SFRs: although the stacking analysis is unable to give such indication (only constraining the average SFR to $<300$\,$M_{\odot}$ yr$^{-1}$), pDRGs are the sources responsible for the observed difference of the average SFR of EROs and that of DRGs in this redshift range (all but 4 of these non-AGN EROs are also classified as DRGs). The pDRGs effectively dilute the stacking signal from 160 (200 if radio detections are included) $M_{\odot}$ yr$^{-1}$ to $110(140)$ $M_{\odot}$ yr$^{-1}$. The latter should thus be taken as a better upper SFR limit than that obtained from the actual pDRG stack.

\begin{figure}
\plotone{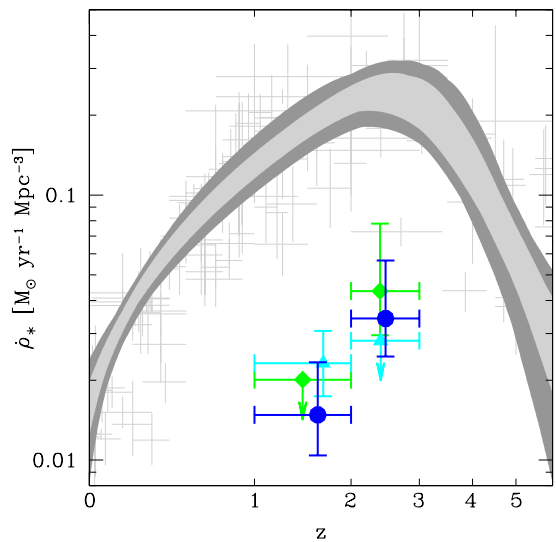}
\caption{Contribution of ERG populations to the total $\dot{\rho}_{\ast}$ for $1\leq{z}<2$ and $2\leq{z}\leq3$. EROs are denoted by green diamonds, IEROs by cyan triangles, and DRGs by blue circles. The compilation of \citet{HopkinsBeacom06} is displayed for reference (gray crosses and shaded region, corresponding to the $\dot{\rho}_{\ast}$ $1\sigma$ and $3\sigma$ confidence regions).\label{sfh}}
\end{figure}

The SFR density behavior for ERGs roughly follows the general trend for SF galaxies, increasing from $1~{\leq}~z<2$ to $2~{\leq}~z<3$ (Figure~\ref{sfh}). Overall, the ERG contribution to the total $\dot{\rho}_{\ast}$ jumps from $\sim10\%$ in the low redshift bin to around 20\% at $2~{\leq}~z\leq3$, where EROs are the highest contributors ($\sim0.043$ $M_{\odot}$ yr$^{-1}$ Mpc$^{-3}$). The exception is the IERO population which apparently, despite being the biggest contributor among the three populations at low-$z$, shows no evolution between the two redshift intervals.

We finally highlight the fact that these estimates may be missing potentially significant SF from the ERGs classified as AGN. The implemented AGN identification criteria were applied in a very conservative way, possibly even removing some non-AGN galaxies from the non-AGN samples. Also, the SFRs associated with the AGN populations are obviously being discarded in this as in similar investigations. This separation is currently necessary given the lack of knowledge on how to remove the AGN contribution to the (radio) emission. The lack of powerful AGN in the ERG populations suggest omitting any attempted AGN exclusion, without significant detriment to the \emph{average} SFR estimates: for example, the entire ERO population in the $2~{\leq}~z\leq3$ redshift range displays an average radio luminosity of $10^{23.51}$\,W\,Hz$^{-1}$, only 10\% higher than that for the non-AGN EROs in the same redshift range\footnote{Note, however, that the increase for IEROs at $1~{\leq}~z<2$ reaches $\gtrsim30\%$.}. However, when no AGN discrimination is attempted, the numerous AGN sources would result in a much more significant increase for the SFR density estimate: in the previous example, the SFR density for the ERO population would rise by $>100\%$, from 0.04 to 0.1 $M_{\odot}$ yr$^{-1}$ Mpc$^{-3}$. The correct scenario will obviously be between the two extremes. A precise handling of such uncertainty can only be fully addressed once a significantly better understanding of the AGN-SF connection is achieved, and more discriminating diagnostics are developed \citep[but see][]{Dunne09}. 

\section{Conclusions} \label{conc}

We have presented a multi-wavelength analysis of the properties of the ERG population in the GOODS-South field. EROs, IEROs, DRGs -- and various combinations between these groups -- are considered, their AGN content identified and their contribution to the global $\dot{\rho}_{\ast}$ estimated, leading to the following conclusions. 

\begin{itemize}

\item The different criteria for the selection of red galaxies select, as previously known, sources at different redshift ranges: while the bulk of EROs and IEROs can be found at $1<z<2$, DRGs are mostly found at $2<z<3$. Different combinations of the three criteria result in samples with distinct redshift properties: while cERGs are observed in a wide redshift range, $1<z<3$, and have no low-$z$ ($z<1$) interlopers, pEROs and pDRGs appear in distinct redshift intervals, at $1<z<2$ and $2<z<4$, respectively. The ``pure'' criteria appear, thus, to be suitable and simple techniques to select high-z sources in well constrained redshift intervals. See Section \ref{reddis}.

\item the ERG population does not include a large number of {\emph {powerful}} AGN, as indicated by the X-rays and radio observations. Almost one-third of the ERG sample hosts potential AGN activity, with the fraction of AGN increasing from EROs to IEROs to DRGs (respectively, 29\%, 35\%, and 38\%). Among ERGs, and according to the X-ray properties, type-2 sources dominate (a 2:1 ratio). An X-ray estimate of the type-2 to type-1 AGN ratio among the ERG population is, however, indeterminate, requiring observations extending to lower X-ray energies (higher wavelengths). See Section \ref{xr} and Section \ref{agncont}.

\item EROs at $z<2$ are often pEROs (56\%), which are likely passively evolved systems without strong SFR activity, on average below $\sim20$ $M_{\odot}$ yr$^{-1}$. On the other hand, essentially all EROs at $2<z<3$ are classified as DRGs and have intense SF activity, similar to ULIRGs (on average $\sim200$ $M_{\odot}$ yr$^{-1}$). DRGs also show this increase in average SFR from low ($1<z<2$) to high ($2<z<3$) redshifts, although not as pronounced (from $\sim70$ to $\sim140$ $M_{\odot}$ yr$^{-1}$). See Section \ref{sfrfin}.

\item The overlapping population between EROs and DRGs displays an intense average SFR both at $1\leq{z}<2$ \citep[$\sim70$ $M_{\odot}$ yr$^{-1}$, supporting previous claims of a dusty starburst nature for these sources,][]{Smail02,Papovich06} and $2\leq{z}\leq3$ ($\sim200$ $M_{\odot}$ yr$^{-1}$). See Section \ref{sfrfin}.

\item The contribution of ERGs to the SFR density increases with redshift: from $\sim10\%$ at $1<z<2$ to $\sim20\%$ at $2<z<3$, an increase which is attributable to the ERO and DRG populations. IEROs, on the other hand, despite showing the highest contribution to the global SF history among the three ERG population at low-$z$, appear to show no evolution at high redshifts. See Section \ref{sfrfin}.

\item SFR densities from ERG populations were estimated for SF-dominated sources, after a thorough AGN multi-wavelength identification. We find that, although the AGN ERGs would only slightly increase the average radio luminosities shown by the non-AGN samples, inclusion of such sources in the SFR density estimates could lead to significant (and, at this point, undetermined) biases. See Section \ref{sfrfin}.

\item The use of a [5.8]-[8.0] versus [8.0]-[24] color diagnostic allows for a tentative separation between AGN and SF galaxies at $z>2.5$, where mid-IR (3 to 8\,$\mu$m) diagnostics become degenerate. In particular, the use of this diagnostic enables the identification of a $z\sim2.5$ evolved system candidate with MIR colors typical of Sc galaxies. See Section \ref{mir25}.

\end{itemize}

\acknowledgments
We thank the FIREWORKS team for the catalog publicly supplied \citep{Wuyts08}; Tommy Wiklind, Du\'{i}lia de Mello, Leonidas Moustakas, Tomas Dahlen, Harry Ferguson, Norman Grogin, Andrea Comastri, and Jennifer Donley for discussion and comments, and Vernesa Smol\v{c}i\'{c} for the talk invitation at California Institute of Technology. HM acknowledges support from Funda\c{c}\~{a}o para a Ci\^{e}ncia e a Tecnologia (FCT, Portugal) through the scholarship SFRH/BD/31338/2006, and from the Anglo-Australian Observatory during a visit to its headquarters. HM and JA acknowledge support from FCT through the research grant PTDC/FIS/100170/2008. AMH acknowledges support provided by the Australian Research Council through a QEII Fellowship (DP0557850). TD acknowledges support from FCT (POCI/CTE-AST/58027/2004). DMA thanks the Royal Society and Leverhulme Trust for financial support.

\clearpage
\begin{deluxetable}{cr|rrr|r|rrrrrr|rrr|r|r|r}
\tabletypesize{\scriptsize}
\rotate
\tablecaption{ERG Number Statistics: Sample Overlap, Counterparts, and Classification.\label{tabctp}}
\tablewidth{0pt}
\tablehead{
\colhead{POP} & \colhead{$N_{\rm{TOT}}$\tablenotemark{a}} & \colhead{\tiny{ERO}} & \colhead{\tiny{IERO}} & \colhead{\tiny{DRG}} & \colhead{$N_{\rm{spec}}$} & \multicolumn{6}{c}{X-Ray} & \colhead{IRAC\tablenotemark{d}} & \colhead{PL\tablenotemark{e}} & \colhead{MIR cc\tablenotemark{e}} & \colhead{MIPS} & \colhead{Radio} & \colhead{N\tiny{$_{\rm{AGN}}$}\tablenotemark{f}} \\
\colhead{} & \colhead{} & \colhead{} & \colhead{} & \colhead{} & \colhead{} & \colhead{\tiny{XR}\tablenotemark{b}} & \colhead{\tiny{A1}\tablenotemark{c}} & \colhead{\tiny{A2}\tablenotemark{c}} & \colhead{\tiny{Q1}\tablenotemark{c}} & \colhead{\tiny{Q2}\tablenotemark{c}} & \colhead{\tiny{nHR}\tablenotemark{c}} & \colhead{} & \colhead{} & \colhead{} & \colhead{$24\mu$m} & \colhead{\tiny{$1.4\,\rm{GHz}$}} & \colhead{}
}
\startdata
\bf{ERG} & 731 (702) & 645 & 294 & 350 & 155 & 88 & 18 & 36 & 1 & 5 & 26 & 675 & 14 (1) & 164 (30) & 402 & 17 & 214 (29\%) \\ 
\bf{ERO} & 645 (626) & 645 & 290 & 267 & 145 & 84 & 17 & 35 & 1 & 4 & 25 & 612 & 6 (1) & 135 (27) & 359 & 15 & 184 (29\%) \\
\bf{IERO} & 294 (289) & 290 & 294 & 198 & 29 & 47 & 9 & 19 & 1 & 3 & 14 & 290 & 4 (1) & 83 (21) & 191 & 10 & 103 (35\%) \\
\bf{DRG} & 350 (338) & 267 & 198 & 350 & 39 & 51 & 11 & 17 & 0 & 5 & 16 & 322 & 12 (0) & 117 (25) & 222 & 11 & 135 (39\%) \\
\bf{cERG} & 197 (197) & 197 & 197 & 197 & 17 & 36 & 8 & 14 & 0 & 3 & 10 & 197 & 2 (0) & 69 (18) & 141 & 7 & 81 (41\%) \\
\bf{pERO} & 272 (272) & 272 & 0 & 0 & 103 & 24 & 6 & 13 & 0 & 0 & 5 & 257 & 0 (0) & 34 (2) & 128 & 3 & 56 (21\%) \\
\bf{pDRG} & 70 (70) & 0 & 0 & 70 & 9 & 4 & 1 & 1 & 0 & 1 & 1 & 51 & 6 (0) & 24 (3) & 35 & 2 & 25 (36\%) \\
\enddata
\tablenotetext{a}{Total number of sources in each (sub)sample and, in parenthesis, those which have good photometry in all bands involved in the ERG criteria: $i_{775}$, $z_{850}$, $J$, $K_s$, and $3.6\mu$m.}
\tablenotetext{b}{Total number of X-ray identifications.}
\tablenotetext{c}{Number of sources classified as type-1 or type-2 AGN (A1 or A2, respectively), type-1 or type-2 QSO (Q1 or Q2, respectively), and AGN with undetermined type (no HR determination, column nHR) according to the \citet{Szokoly04} criterion.}
\tablenotetext{d}{Number of sources with detection in all four IRAC channels.}
\tablenotetext{e}{Sources classified as AGN by PL \citep{Donley07} or MIR color--color \citep{Stern05} criteria. In parenthesis, those also classified as AGN by the X-ray criterion.}
\tablenotetext{f}{Total number of sources classified as AGN, considering all AGN identification criteria, along with the equivalent fraction in the total (sub)population.}
\end{deluxetable}
\clearpage

\begin{deluxetable}{crrrr}
\tabletypesize{\scriptsize}
\tablecaption{Robust Radio stacking of ERG populations.\label{tabnum}}
\tablewidth{0pt}
\tablehead{
\colhead{POP} & \colhead{$N_{\rm{TOT}}$} & 
\colhead{$N_{<18''}$\tablenotemark{a}} & \colhead{$N_{3\sigma}$\tablenotemark{b}} & \colhead{$N_{\rm{fin}}$\tablenotemark{c}}
}
\startdata
\bf{EROs} \\  
$z12$ &  443 & 56 & 12 & 370 \\ 
$z12;\rm{nAGN}$ &  344 & 37 & 7 & 296 \\ 
$z23$ &  144 & 16 & 3 & 124 \\ 
$z23;\rm{nAGN}$ &  84 & 6 & 2 & 75 \\ 
\bf{IEROs} \\  
$z12$ &  161 & 20 & 6 & 132 \\ 
$z12;\rm{nAGN}$ &  124 & 11 & 5 & 106 \\ 
$z23$ &  111 & 12 & 2 & 96 \\ 
$z23;\rm{nAGN}$ &  59 & 3 & 1 & 54 \\ 
\bf{DRGs} \\  
$z12$ &  102 & 9 & 3 & 88 \\ 
$z12;\rm{nAGN}$ &  71 & 4 & 2 & 64 \\ 
$z23$ &  176 & 20 & 3 & 152 \\ 
$z23;\rm{nAGN}$ &  111 & 7 & 2 & 101 \\ 
\bf{cERGs} \\  
$z12$ &  71 & 7 & 2 & 60 \\ 
$z12;\rm{nAGN}$ &  49 & 3 & 2 & 43 \\ 
$z23$ &  104 & 12 & 2 & 89 \\ 
$z23;\rm{nAGN}$ &  59 & 3 & 1 & 54 \\ 
\bf{pEROs} \\  
$z12$ &  245 & 34 & 5 & 204 \\ 
$z12;\rm{nAGN}$ &  193 & 25 & 2 & 164 \\ 
$z23$ &  5 & 1 & 0 & 4 \\ 
$z23;\rm{nAGN}$ &  4 & 1 & 0 & 3 \\ 
\bf{pDRGs} \\  
$z12$ &  1 & 0 & 0 & 0 \\
$z12;\rm{nAGN}$ &  1 & 0 & 0 & 0 \\ 
$z23$ &  34 & 4 & 0 & 30 \\ 
$z23;\rm{nAGN}$ &  25 & 1 & 0 & 24 \\ 
\enddata
\tablecomments{The $z12$ and $z23$ abbreviations stand for $1\leq{z}<2$ and $2\leq{z}\leq3$, respectively.}
\tablenotetext{a}{Number of stamps with a radio detection within 18'' of the ERG position, consequently rejected from the final stacking.}
\tablenotetext{b}{Number of stamps with a possible radio detection at the ERG position (signal between 3$\sigma$ and $\sim 4.5\sigma$), also removed from the final stacking.}
\tablenotetext{c}{Final number of stamps included in the stacking.}
\end{deluxetable}

\clearpage
\begin{deluxetable}{crrrrrrrrrrrr}
\tabletypesize{\scriptsize}
\rotate
\tablecaption{Properties of Extremely Red Galaxy Populations from Radio Stacking Analysis\label{tabsta}}
\tablewidth{0pt}
\tablehead{
\colhead{POP} & \colhead{$N_{\rm{TOT}}$} & \colhead{$N_{\rm{fin}}$\tablenotemark{a}} & \colhead{rms} & \colhead{S$_{\rm 1.4GHz}$} & \colhead{S/N} & \colhead{$N_{MC}$\tablenotemark{b}} & \colhead{$\overline{z}$} & \colhead{$\log$ ($L_{1.4\,{\rm GHz}}$)\tablenotemark{c}} & \colhead{SFR\tablenotemark{c}} & \colhead{$\log$ ($\mathcal{L}_{1.4\,{\rm GHz}}$)\tablenotemark{c}} & \colhead{$\dot{\rho}_{\ast}$\tablenotemark{c}}\\
\colhead{} & \colhead{} & \colhead{} & \colhead{($\mu$Jy)} & \colhead{($\mu$Jy)} & \colhead{} & \colhead{} & \colhead{} & \colhead{(W$\,$Hz$^{-1}$)} & \colhead{($M_{\odot}\,$yr$^{-1}$)} & \colhead{(W$\,$Hz$^{-1}\,$Mpc$^{-3}$)} & \colhead{($M_{\odot}\,$yr$^{-1}\,$Mpc$^{-3}$)}
}
\startdata
\bf{EROs} \\   
$z12$ &  443 & 370 & 0.927 & 2.051 & 2.212 & 0 & 1.45 & 22.37(22.83) & \ldots & 19.41(19.90) & \ldots \\ 
$z12;\rm{nAGN}$ &  344 & 296 & 1.037 & 1.080 & 1.041 & 5 & 1.45 & $<$22.25(22.62) & $<$10(23) & $<$19.19(19.56) & $<$8.48e-03(2.01e-02) \\ 
$z23$ &  144 & 124 & 1.708 & 8.641 & 5.059 & 0 & 2.44 & 23.51(23.62) & \ldots & 20.05(20.17) & \ldots \\ 
$z23;\rm{nAGN}$ &  84 & 75 & 2.006 & 8.274 & 4.125 & 0 & 2.39 & 23.47(23.55) & 162(196) & 19.81(19.90) & 3.60e-02(4.34e-02) \\ 
\bf{IEROs} \\  
$z12$ &  161 & 132 & 1.627 & 6.700 & 4.118 & 0 & 1.66 & 23.01(23.25) & \ldots & 19.63(19.96) & \ldots \\ 
$z12;\rm{nAGN}$ &  124 & 106 & 1.792 & 5.071 & 2.830 & 0 & 1.67 & 22.89(23.08) & 43(66) & 19.42(19.62) & 1.45e-02(2.32e-02) \\ 
$z23$ &  111 & 96 & 1.928 & 8.176 & 4.241 & 0 & 2.47 & 23.50(23.58) & \ldots & 19.94(20.03) & \ldots \\ 
$z23;\rm{nAGN}$ &  59 & 54 & 2.369 & 7.103 & 2.998 & 1 & 2.40 & $<$23.46(23.49) & $<$161(170) & $<$19.68(19.71) & $<$2.67e-02(2.82e-02) \\ 
\bf{DRGs} \\  
$z12$ &  102 & 88 & 2.001 & 6.823 & 3.410 & 0 & 1.59 & 22.99(23.12) & \ldots & 19.39(19.56) & \ldots \\ 
$z12;\rm{nAGN}$ &  71 & 64 & 2.283 & 6.711 & 2.940 & 0 & 1.61 & 22.99(23.09) & 54(68) & 19.27(19.43) & 1.04e-02(1.48e-02) \\ 
$z23$ &  176 & 152 & 1.596 & 6.144 & 3.850 & 0 & 2.51 & 23.39(23.52) & \ldots & 19.95(20.10) & \ldots \\ 
$z23;\rm{nAGN}$ &  111 & 101 & 1.744 & 5.224 & 2.995 & 0 & 2.47 & 23.30(23.40) & 110(137) & 19.69(19.79) & 2.73e-02(3.42e-02) \\ 
\bf{cERGs} \\  
$z12$ &  71 & 60 & 2.542 & 8.901 & 3.502 & 0 & 1.67 & 23.15(23.26) & \ldots & 19.49(19.64) & \ldots \\ 
$z12;\rm{nAGN}$ &  49 & 43 & 2.982 & 7.275 & 2.440 & 6 & 1.71 & $<$23.16(23.26) & $<$80(100) & $<$19.39(19.55) & $<$1.36e-02(1.96e-02) \\ 
$z23$ &  104 & 89 & 1.961 & 7.270 & 3.707 & 0 & 2.48 & 23.45(23.55) & \ldots & 19.86(19.97) & \ldots \\ 
$z23;\rm{nAGN}$ &  59 & 54 & 2.369 & 7.103 & 2.998 & 1 & 2.40 & $<$23.46(23.49) & $<$161(170) & $<$19.68(19.71) & $<$2.67e-02(2.82e-02) \\  
\bf{pEROs} \\
$z12$ &  245 & 204 & 1.289 & -0.449 & -0.348 & 1347 & 1.33 & $<$22.36(22.66) & \ldots & $<$19.17(19.47) & \ldots \\ 
$z12;\rm{nAGN}$ &  193 & 164 & 1.413 & -1.036 & -0.733 & 3764 & 1.32 & $<$22.52(22.68) & $<$18(26) & $<$19.24(19.39) & $<$9.54e-03(1.36e-02) \\ 
$z23$ &  6 & \ldots & \ldots & \ldots & \ldots & \ldots & \ldots & \ldots & \ldots & \ldots & \ldots \\ 
$z23;\rm{nAGN}$ &  4 & \ldots & \ldots & \ldots & \ldots & \ldots & \ldots & \ldots & \ldots & \ldots & \ldots \\ 
\bf{pDRGs} \\  
$z12$ &   1 & \ldots & \ldots & \ldots & \ldots & \ldots & \ldots & \ldots & \ldots & \ldots & \ldots \\
$z12;\rm{nAGN}$ &   1 & \ldots & \ldots & \ldots & \ldots & \ldots & \ldots & \ldots & \ldots & \ldots & \ldots \\
$z23$ &  34 & 30 & 3.713 & 1.049 & 0.283 & 3396 & 2.67 & $<$23.71(23.75) & \ldots & $<$19.67(19.72) & \ldots \\ 
$z23;\rm{nAGN}$ &  25 & 24 & 4.037 & -1.532 & -0.379 & 7001 & 2.65 & $<$23.80(23.80) & $<$345(345) & $<$19.70(19.70) & $<$2.77e-02(2.77e-02) \\ 
\enddata
\tablecomments{The $z12$ and $z23$ abbreviations stand for $1\leq{z}<2$ and $2\leq{z}\leq3$, respectively. The upper limits for Luminosity and SFR estimates, whenever $N_{MC}>0$, are calculated considering the maximum S/N obtained in the respective set of MC simulations.}
\tablenotetext{a}{Final number of stamps included in the stacking, after the various rejection steps described in Section \ref{stack}.}
\tablenotetext{b}{Number of MC simulations (out of 10000) that resulted in higher S/N values.}
\tablenotetext{c}{In parenthesis, the estimated value also takes into account radio detections ($>3\sigma$) excluded from the stacking procedure (see Section \ref{stack}).}
\end{deluxetable}
\clearpage

\end{document}